\documentclass[aps,floatfix,prb,twocolumn,a4paper,10pt,showpacs,citeautoscript,superscriptaddress,preprintnumbers]{revtex4-1} 
\usepackage[english]{babel}		

\usepackage{amsmath}		
\usepackage{amssymb}		
\usepackage{bbm}			

\usepackage{graphicx}		
\usepackage{graphics}		
\DeclareGraphicsExtensions{.png,.pdf}
\usepackage{xcolor}		
\usepackage[colorlinks=true,a4paper=true, pdfstartview=FitV,linkcolor=blue, citecolor=blue,urlcolor=blue]{hyperref}

\renewcommand{\theequation}{\arabic{equation}}

\newcommand{\Equation}[2]{\begin{equation}\label{#1}#2\end{equation}}
\newcommand{\Align}[2]{\begin{align}\label{#1}#2\end{align}}

\newcommand{\bs}{\boldsymbol}

\newcommand{\Figref}[1]{Fig.~\ref{#1}}
\newcommand{\Eqref}[1]{\eqref{#1}}

\newcommand{\eg}{{\it e.g.~}}
 
\newcommand{\ie}{{\it i.e.~}} 
\newcommand{\resp}{{\it resp.~}} 

\newcommand{\Uone}{\mbox{U(1)}}
\newcommand{\Ztwo}{\mathbbm{Z}_2}
\newcommand{\Relative}{\mathbbm{Z}}
\newcommand{\Real}{\mathbbm{R}}

\newcommand{\dd}{\mathrm{d}}
\newcommand{\Exp}[1]{\mathrm{e}^{#1}}
\renewcommand\Re{\mathrm{Re}}
\renewcommand\Im{\mathrm{Im}}

\newcommand{\oo}{{(1)}}
\newcommand{\ot}{{(2)}}
\newcommand{\op}{{(+)}}
\newcommand{\om}{{(-)}}
\newcommand{\opm}{{(\pm)}}

\newcommand{\SRO}{Sr$_2$RuO$_{4}$~}
\newcommand{\Q}{\mathcal{Q}}

\begin{document}
\preprint{Phys. Rev. B. {\bf XX}, XXXXXX (2012)}
\title{\texorpdfstring{Skyrmionic state and stable half-quantum vortices in chiral $p$-wave superconductors}
{Skyrmionic state and stable half-quantum vortices in chiral p-wave superconductors}}
\author{Julien~Garaud}
\author{Egor~Babaev}
\affiliation{Department of Physics, University of Massachusetts Amherst, MA 01003 USA}
\affiliation{Department of Theoretical Physics, The Royal Institute of Technology, Stockholm, SE-10691 Sweden}
\begin{abstract}
Observability of half-quantum vortices and skyrmions in $p$-wave superconductors is an 
outstanding open question. Under the most common conditions, fractional flux vortices 
are not thermodynamically stable in bulk samples. Here we show that in chiral  
$p$-wave superconductors,  there is a regime where, in contrast  lattices of integer flux 
vortices are not thermodynamically stable. Instead skyrmions made of spatially separated 
half-quantum vortices are the topological defects produced by an applied external field. 
\end{abstract}

\maketitle

Higher broken symmetries in $p$-wave superconductors have inspired long-standing interest 
to realize topological defects more complicated than vortices. Much of the early discussions 
of various complex topological defects were in the context of superfluid ${}^3 \mathrm{He}$ 
\cite{volovik,*Anderson.Toulouse:77,*Thuneberg:86,*Salomaa.Volovik:86,
*Salomaa.Volovik:87,*Tokuyasu.Hess.ea:90}. Recently attention to these questions has
raised dramatically in connection with superconductors which are argued to have $p$-wave 
pairing, such as \SRO. The highly interesting possibility there, is connected with half-quantum 
vortices \cite{science,spinsf,sukbum1,sukbum2,sukbum3,frac3,frac}. Their statistics is 
non-Abelian and they could potentially be used for quantum computations \cite{kitaev}. 
Other kinds of topological defects discussed in connection with spin-triplet superconductors 
are skyrmions \cite{Knigavko.Rosenstein:99,*Knigavko.Rosenstein.ea:99,*Rosenstein.Shapiro.ea:03,
Li.Toner.ea:09,*Li.Toner.ea:07} and hopfions \cite{hopfions}. 
In superconducting materials, the creation of these topological excitations is  highly nontrivial.
Superconducting components are coupled by a gauge field and there are also 
symmetry-reducing inter-component interactions. As a consequence fractional vortices 
have logarithmically or linearly divergent energies (see \eg Ref.~\onlinecite{frac}), while 
integer flux vortices have finite energy per unit length. Consequently, under usual conditions, 
half-quantum vortices are thermodynamically unstable in bulk systems. It was argued that 
complex setups, such as mesoscopic samples, are needed for their creation \cite{frac,science,
chibotaru,*chibotaru2,*bluhm,*geurts,*pina}. Recently it was claimed that a half-quantum 
vortex was observed in a mesoscopic sample of \SRO \cite{science}. Other proposed routes 
to observe fractional vortices, invoke (i) thermal deconfinement \cite{npb,*frac2,spinsf,sukbum3},  
(ii) potential materials with strongly reduced spin stiffness \cite{sukbum1} and (iii) regimes 
very close to the upper critical magnetic field, where gauge-field mediated half-quantum vortex 
confinement is weak \cite{sukbum2}. 
In some more general systems it was shown that fractional vortices could be thermodynamically 
stable near boundaries \cite{silaev}. 
Today the conditions under which half-quantum vortices and skyrmions 
\cite{Knigavko.Rosenstein:99,*Knigavko.Rosenstein.ea:99,*Rosenstein.Shapiro.ea:03,
Li.Toner.ea:09,*Li.Toner.ea:07} could be experimentally created in bulk superconductors 
still remains an outstanding open question.

\begin{figure}[!htb]
 \hbox to \linewidth{ \hss
 \includegraphics[width=\linewidth]{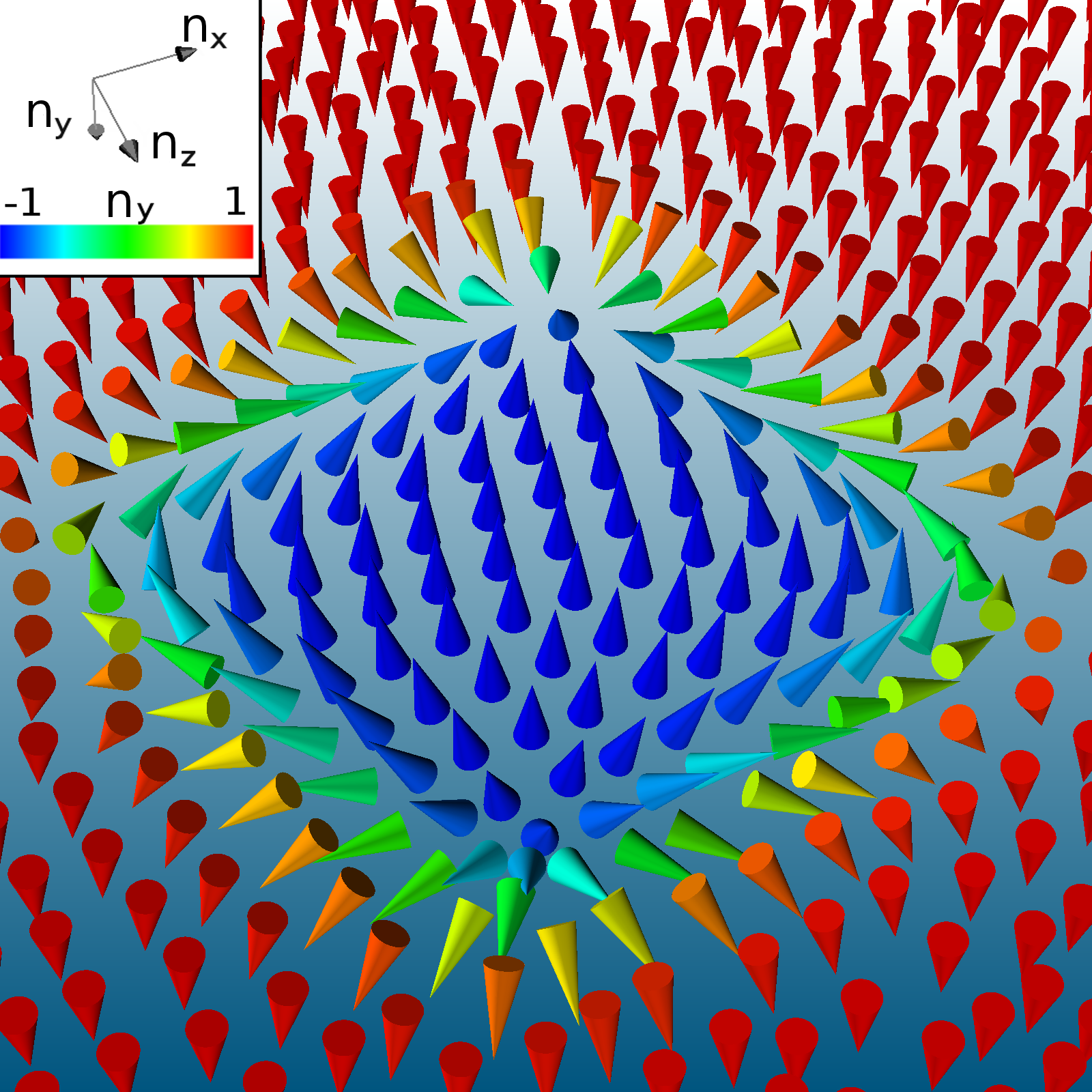}
 \hss}
\caption{
(Color on-line) -- 
Numerically calculated texture of the pseudo-spin vector for a skyrmion carrying 
with a topological charge $\Q=2$. As can be seen in the picture the skyrmionic 
topological charge density is confined in a closed domain wall.
}
\label{Fig:Single-texture}
\end{figure}

In this work we investigate the magnetic response of the Ginzburg-Landau model the has 
been widely applied to \SRO \cite{Agterberg:98,Agterberg:98a,*Heeb.Agterberg:99}. Our 
considerations apply to two-dimensional systems or three-dimensional problems with 
translation invariance along the $z$-direction. Then the free energy density reads 
\begin{figure}[!htb]
  \hbox to \linewidth{ \hss
  \includegraphics[width=\linewidth]{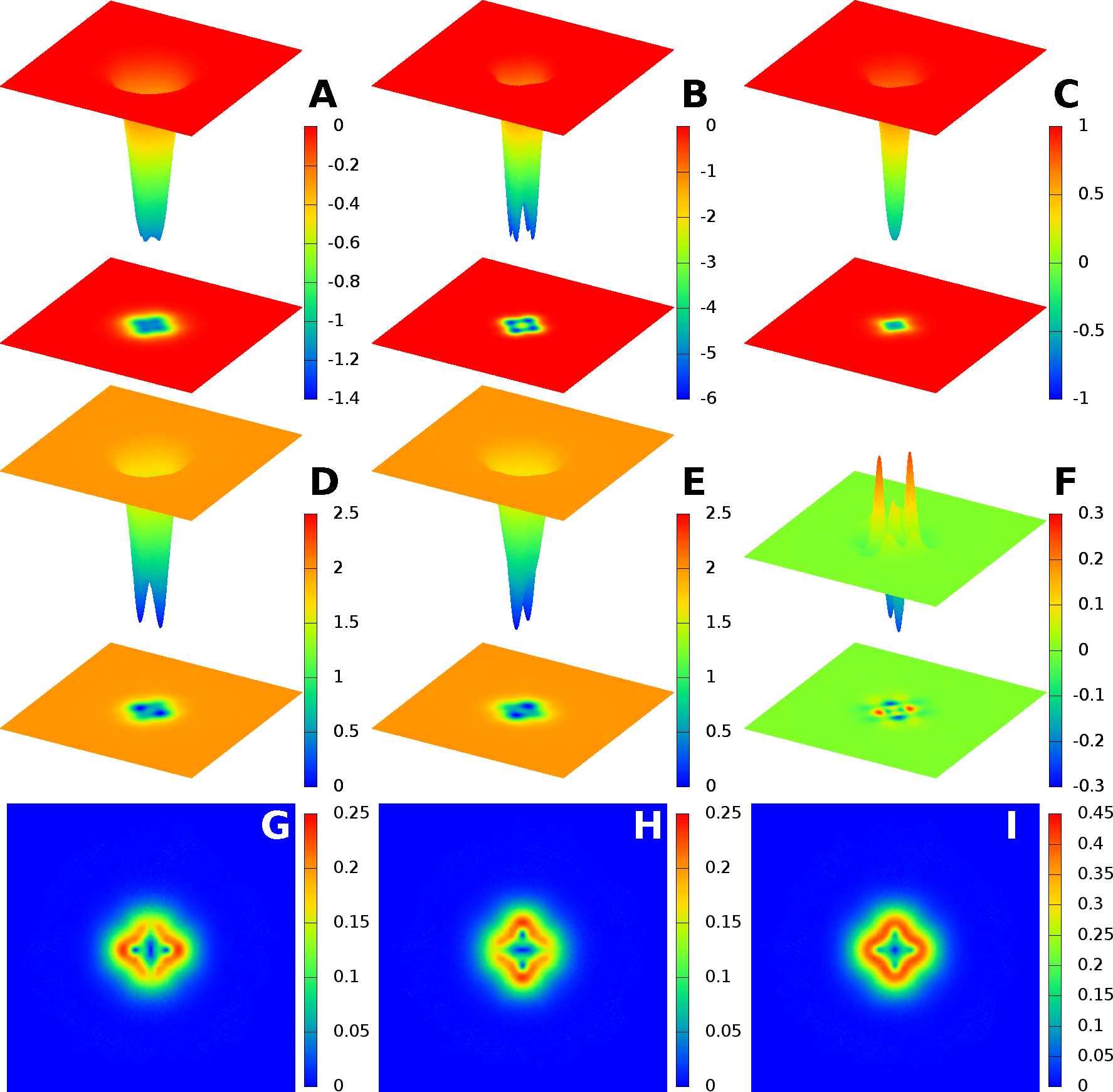}
 \hss}
\caption{
(Color on-line) -- 
A thermodynamically stable skyrmion carrying two flux quanta, with $e=0.8$ and 
$\gamma=0.5$. 
Displayed quantities are, magnetic flux ($\bf A$), the (inverted) energy density ($\bf B$) 
and  the sine of the phase difference $\sin(\varphi_{2}-\varphi_{1})$ ($\bf C$). On the 
second line, the densities of superconducting order parameter components  
$|\psi_1|^2$ ($\bf D$), $|\psi_2|^2$ ($\bf E$), and the `doubled phase difference' 
$\Im(\psi_1^{*\ 2}\psi_2^2)$ ($\bf F$). Panels ($\bf G$) and (\resp $\bf H$) on the third
line are the supercurrents associated with each component $\psi_1$ (\resp $\psi_2$) of 
the order parameter (see Appendix \ref{Appendix} for definition). The last panel ($\bf I$) 
shows the total supercurrent.
}
\label{Fig:Skyrmion1}
\end{figure}
\begin{subequations}\label{freeEnergy}
\begin{align}
\mathcal{F}&(\psi_a,\bs A)= |\nabla\times\bs A|^2  \label{magneticEnergy}\\
   &+|D_x\psi_1|^2+ |D_y\psi_2|^2+\gamma |D_y\psi_1|^2+ \gamma| D_x\psi_2|^2	\nonumber \\
   &+2\gamma\Re\left[ (D_x\psi_1)^*D_y\psi_2+ (D_y\psi_1)^*D_x\psi_2 \right]	 \label{gradientEnergy}\\
   &+(2\gamma-1)|\psi_1|^2|\psi_2|^2+\sum_{a=1,2}-|\psi_a|^2+\frac{1}{2}|\psi_a|^4
\label{potentialEnergy} \\
   &+\gamma|\psi_1|^2|\psi_2|^2\cos(2(\varphi_2-\varphi_1))	\,.\label{breakEnergy}
\end{align}
\end{subequations}
The different components of the order parameter are denoted 
$\psi_{1,2}=|\psi_{1,2}|\Exp{i\varphi_{1,2}}$; $\bs D=\nabla+ie \bs A$. The 
$p$-wave state is described here by a doublet of complex  fields subjected to
the the following symmetry breaking coupling :
$\Re\left(\psi_1^{*\,2}\psi_2^2\right)=|\psi_1|^2|\psi_2|^2\cos(2(\varphi_2-\varphi_1))$. 
The ground state breaks the $U(1)\times\Ztwo$ symmetry, since the ground state phase 
difference is either $\pi/2$ or $3\pi/2$. Gradient terms \Eqref{gradientEnergy} make this 
model clearly anisotropic in the $xy$-plane. The coefficient $\gamma$, controlling the 
anisotropy, should be $\gamma>1/3$ for \SRO, according to 
\cite{Agterberg:98}. The coupling constant $e$ is a  convenient quantity to parametrize 
the penetration depth of the magnetic field. The discrete $\Ztwo$ symmetry dictates that 
the system allows domain wall solutions interpolating between two regions with different 
phase-locking. Such domain walls are energetically expensive and thus not intrinsically 
stable. It was suggested that they  could be observable if pinned by crystalline defects 
\cite{sigrist,goldbart,*vakaryuk}. Also domain walls formed as dynamic excitations inside 
vortex lattices were studies extensively in \cite{machida1,*machida2,*machida3}. They 
could be experimentally observable in these setups since they  pin half-quantum vortices 
\cite{sigrist,machida1,*machida2,*machida3,*goldbart}.

Returning to the discussion of vortices one can observe that  the system \Eqref{freeEnergy} 
has $\Uone\times\Ztwo$ broken symmetry. Thus a single half-quantum vortex (with winding 
only in one of the phases) has linearly diverging energy and thus is not thermodynamically 
stable \cite{frac}. Also from this broken symmetry, the existence 
of skyrmionic excitations would not follow. The previous works required higher broken symmetry 
for the existence of skyrmions \cite{Knigavko.Rosenstein:99,*Knigavko.Rosenstein.ea:99,
*Rosenstein.Shapiro.ea:03,Li.Toner.ea:09,*Li.Toner.ea:07}. However we show below that there 
is a considerable window of parameters where the system \Eqref{freeEnergy} possesses what 
we term as a ``skyrmionic phase". In that phase, mostly because of favorable competition between 
field gradients and potential and magnetic energies, the system does have \emph{thermodynamically 
stable} skyrmions while ordinary integer flux vortex lattices are \emph{not} thermodynamically 
stable. These skyrmions are bound states of spatially separated half-quantum vortices, connected 
by domain walls. Half-quantum vortices are linearly confined into integer vortices in a bulk sample 
because of the terms $|\psi_1|^2|\psi_2|^2\cos(2(\varphi_2-\varphi_1))$. However on a (closed) 
domain wall, a composite vortex should  split along this wall, since the above-mentioned term 
has there, unfavorable values of the phase difference. Indeed, such deconfining allows to reduce 
energetically unfavorable values of the phase differences. Because of this vortex splitting and 
resulting repulsive interactions, vortices trapped on domain wall can prevent the collapse of a 
closed domain wall. The main result of this paper is that we show that these objects are 
characterized by an integer-valued skyrmionic topological charge and that they can be energetically 
cheaper than vortices. Such a skyrmion is displayed in \Figref{Fig:Single-texture}, as a texture of 
a pseudo-spin vector field defined later on. 

\begin{figure}[!htb]
  \hbox to \linewidth{ \hss
 \includegraphics[width=\linewidth]{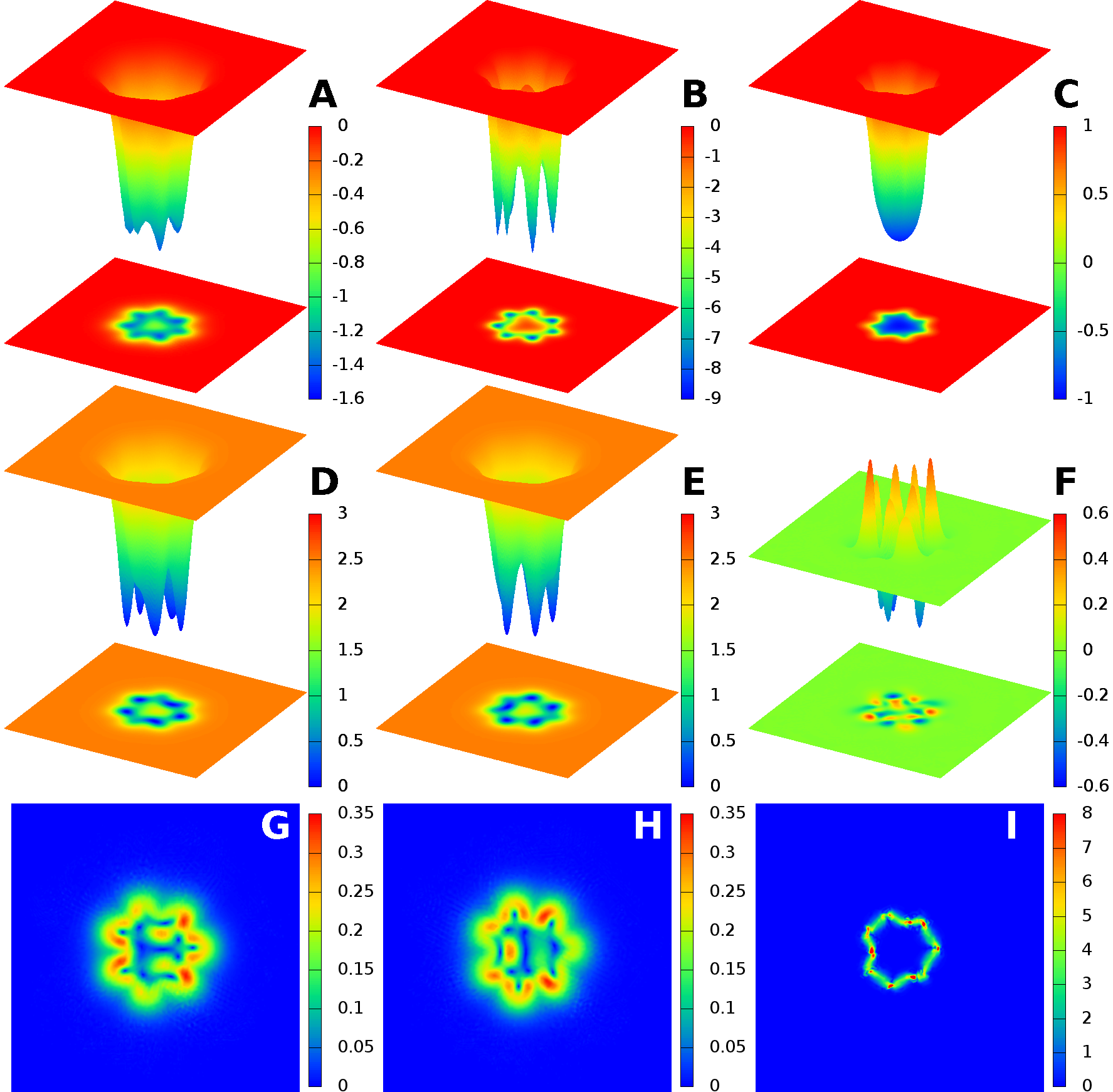}
 \hss}
\caption{
(Color on-line) -- 
A skyrmion carrying five flux quanta, with $e=0.8$ and $\gamma=0.4$. Displayed quantities 
are the same as in \Figref{Fig:Skyrmion1}, except panel ($\bf I$) showing the gradient of the 
phase difference $\nabla \varphi_{12}$, which is non zero at the domain wall. The skyrmion 
consists of ten spatially separated half-quantum  vortices. It assumes a complicated non-symmetric 
structure due to a competition of a preferred geometry of a skyrmion with the anisotropies 
\Eqref{gradientEnergy}.
}
\label{Fig:Skyrmion2}
\end{figure}

We investigated structures carrying $N$ flux quanta (\ie with each phase winding 
$\oint\nabla\varphi_a=2\pi N$) as functions of the gauge coupling $e$ and the anisotropy 
parameter $\gamma$.  Ground states, carrying a given number of magnetic flux quanta, 
are computed numerically by minimizing the energy within a finite element framework 
provided by the Freefem++ library \cite{Freefem}. See technical details in Appendix 
\ref{Numerics}. 

\begin{figure}[!htb]
  \hbox to \linewidth{ \hss
  \includegraphics[width=\linewidth]{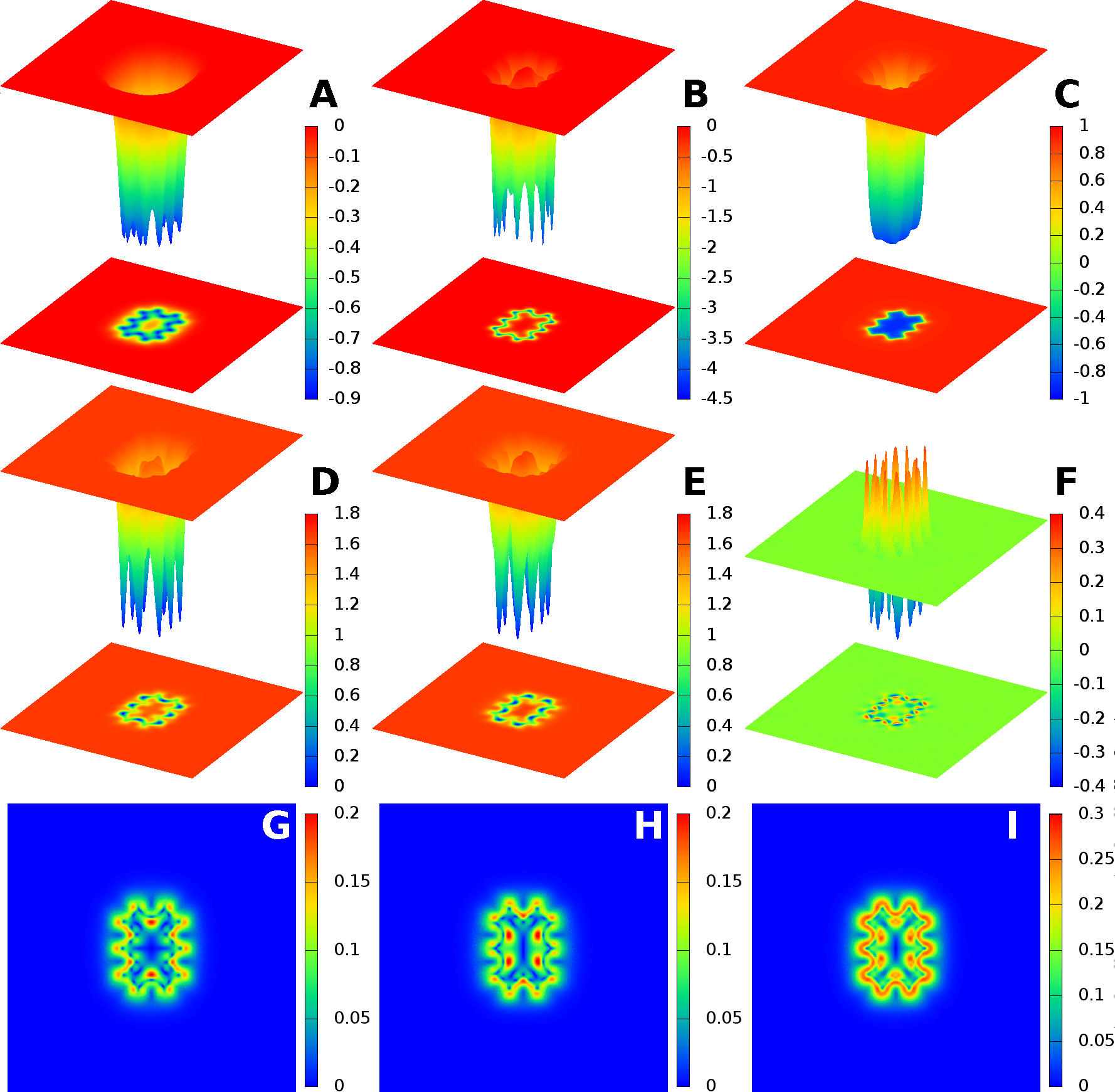}
 \hss}
\caption{
(Color on-line) -- 
A skyrmion with $N=8$, $e=0.6$ and in the case of higher anisotropy $\gamma=0.6$.
Displayed quantities are the same as in \Figref{Fig:Skyrmion1}.
}
\label{Fig:Skyrmion3}
\end{figure}

We find that if the penetration length is sufficiently large  (\ie at small values of the coupling 
constant $e$), the system indeed forms ordinary Abrikosov vortices in external field. 
On the other hand for sufficiently large $e$ the system behaves as a type-I superconductor. 
However there is a regime  in a wide range of intermediate coupling constants $e$, 
where integer flux vortices are more expensive than bound states of spatially separated 
half-quantum vortices connected by closed domain wall. Such configurations carrying 
different number of flux quanta are given in Figures \ref{Fig:Skyrmion1}, \ref{Fig:Skyrmion2} 
and \ref{Fig:Skyrmion3}. The clearly visible preferred directions for supercurrents originate 
in the anisotropies \Eqref{gradientEnergy}. The cores in different components do not 
coincide in space. This means fractionalization of vortices in this state. Each of the split 
cores carries a half of a flux quantum (for detailed calculations of fractional vortices flux 
quantization, see \eg Ref.~\onlinecite{frac}).

The configurations found here are actually skyrmions, although it may not be obvious 
from the Figures \ref{Fig:Skyrmion1}, \ref{Fig:Skyrmion2} and \ref{Fig:Skyrmion3}. 
To prove that the solutions are skyrmions the two-component model \Eqref{freeEnergy} 
is mapped to an anisotropic non-linear $\sigma$-model \cite{bfn}. In that mapping the 
superconducting condensates are projected on the Pauli matrices $\bs\sigma$ allowing 
to define the pseudo-spin vector $\bf n$:
\Equation{Projection}{
   {\bf n}\equiv (n_x,n_y,n_z)=\frac{\Psi^\dagger\bs \sigma\Psi}{\Psi^\dagger\Psi}
~~~~~\text{where}~~~~~
\Psi^\dagger=(\psi_1^*,\psi_2^*)\, .
}
The target space being a sphere, together with the one-point compactification of the 
plane defines the map ${\bf n}: S^2\to S^2$. Such maps are classified by the homotopy 
class $\pi_2(S^2)\in\Relative$, so there exists an integer valued topological charge  
\Equation{Charge}{
   \Q({\bf n})=\frac{1}{4\pi} \int_{\Real^2}
   {\bf n}\cdot\partial_x {\bf n}\times \partial_y {\bf n}\,\,
  \dd x \dd y \,.
}
{For a skyrmion, $\Q=N$, while $\Q=0$ for ordinary vortices.}
The terms in \Eqref{potentialEnergy} and \Eqref{breakEnergy}, break the $\mathrm{O}(3)$
symmetry of the pseudo-spin $\bf n$ down to $\Ztwo$. In a  non-linear $\sigma$-model, 
such anisotropy would undermine stability of the skyrmions.  However this collapse does 
not occur in the Ginzburg-Landau model, because of the behaviour of the gradient energy, 
which is demonstrated below.

The numerically computed topological charge \Eqref{Charge} is found to be an integer (with 
a negligible relative error of the order $10^{-5}$, due to the discretization) for the closed 
domain wall/vortex systems which are therefore skyrmions. The solutions shown in Figures 
\ref{Fig:Skyrmion1}, \ref{Fig:Skyrmion2} and \ref{Fig:Skyrmion3} have skyrmionic topological 
charge $\Q=2$, $\Q=5$, $\Q=8$ correspondingly. The terminology skyrmion is more intuitively 
obvious when the solutions are represented in terms of the  pseudo-spin vector field ${\bf n}$, 
as in \Figref{Fig:Single-texture}. However unlike skyrmions in non-linear $\sigma$-model, here 
the skyrmionic topological charge density is mostly concentrated on the half-quantum vortices and 
on the domain wall.

\begin{figure}[!htb]
\hbox to \linewidth{ \hss
\begingroup
  \makeatletter
  \providecommand\color[2][]{%
    \GenericError{(gnuplot) \space\space\space\@spaces}{%
      Package color not loaded in conjunction with
      terminal option `colourtext'%
    }{See the gnuplot documentation for explanation.%
    }{Either use 'blacktext' in gnuplot or load the package
      color.sty in LaTeX.}%
    \renewcommand\color[2][]{}%
  }%
  \providecommand\includegraphics[2][]{%
    \GenericError{(gnuplot) \space\space\space\@spaces}{%
      Package graphicx or graphics not loaded%
    }{See the gnuplot documentation for explanation.%
    }{The gnuplot epslatex terminal needs graphicx.sty or graphics.sty.}%
    \renewcommand\includegraphics[2][]{}%
  }%
  \providecommand\rotatebox[2]{#2}%
  \@ifundefined{ifGPcolor}{%
    \newif\ifGPcolor
    \GPcolortrue
  }{}%
  \@ifundefined{ifGPblacktext}{%
    \newif\ifGPblacktext
    \GPblacktexttrue
  }{}%
  \let\gplgaddtomacro\g@addto@macro
  \gdef\gplbacktext{}%
  \gdef\gplfronttext{}%
  \makeatother
  \ifGPblacktext
    \def\colorrgb#1{}%
    \def\colorgray#1{}%
  \else
    \ifGPcolor
      \def\colorrgb#1{\color[rgb]{#1}}%
      \def\colorgray#1{\color[gray]{#1}}%
      \expandafter\def\csname LTw\endcsname{\color{white}}%
      \expandafter\def\csname LTb\endcsname{\color{black}}%
      \expandafter\def\csname LTa\endcsname{\color{black}}%
      \expandafter\def\csname LT0\endcsname{\color[rgb]{1,0,0}}%
      \expandafter\def\csname LT1\endcsname{\color[rgb]{0,1,0}}%
      \expandafter\def\csname LT2\endcsname{\color[rgb]{0,0,1}}%
      \expandafter\def\csname LT3\endcsname{\color[rgb]{1,0,1}}%
      \expandafter\def\csname LT4\endcsname{\color[rgb]{0,1,1}}%
      \expandafter\def\csname LT5\endcsname{\color[rgb]{1,1,0}}%
      \expandafter\def\csname LT6\endcsname{\color[rgb]{0,0,0}}%
      \expandafter\def\csname LT7\endcsname{\color[rgb]{1,0.3,0}}%
      \expandafter\def\csname LT8\endcsname{\color[rgb]{0.5,0.5,0.5}}%
    \else
      \def\colorrgb#1{\color{black}}%
      \def\colorgray#1{\color[gray]{#1}}%
      \expandafter\def\csname LTw\endcsname{\color{white}}%
      \expandafter\def\csname LTb\endcsname{\color{black}}%
      \expandafter\def\csname LTa\endcsname{\color{black}}%
      \expandafter\def\csname LT0\endcsname{\color{black}}%
      \expandafter\def\csname LT1\endcsname{\color{black}}%
      \expandafter\def\csname LT2\endcsname{\color{black}}%
      \expandafter\def\csname LT3\endcsname{\color{black}}%
      \expandafter\def\csname LT4\endcsname{\color{black}}%
      \expandafter\def\csname LT5\endcsname{\color{black}}%
      \expandafter\def\csname LT6\endcsname{\color{black}}%
      \expandafter\def\csname LT7\endcsname{\color{black}}%
      \expandafter\def\csname LT8\endcsname{\color{black}}%
    \fi
  \fi
  \setlength{\unitlength}{0.0500bp}%
\resizebox{115pt}{84pt}{  \begin{picture}(7200.00,5040.00)%
    \gplgaddtomacro\gplbacktext{%
      \csname LTb\endcsname%
      \put(1078,704){\makebox(0,0)[r]{\strut{} \huge 0.88}}%
      \put(1078,1518){\makebox(0,0)[r]{\strut{} \huge 0.92}}%
      \put(1078,2332){\makebox(0,0)[r]{\strut{} \huge 0.96}}%
      \put(1078,3147){\makebox(0,0)[r]{\strut{} \huge 1}}%
      \put(1078,3961){\makebox(0,0)[r]{\strut{} \huge 1.04}}%
      \put(1078,4775){\makebox(0,0)[r]{\strut{} \huge 1.08}}%
      \put(1210,284){\makebox(0,0){\strut{} \huge 1}}%
      \put(2009,284){\makebox(0,0){\strut{} \huge 2}}%
      \put(2808,284){\makebox(0,0){\strut{} \huge 3}}%
      \put(3607,284){\makebox(0,0){\strut{} \huge 4}}%
      \put(4406,284){\makebox(0,0){\strut{} \huge 5}}%
      \put(5205,284){\makebox(0,0){\strut{} \huge 6}}%
      \put(6004,284){\makebox(0,0){\strut{} \huge 7}}%
      \put(6803,284){\makebox(0,0){\strut{} \huge 8}}%
      \put(100,2739){\rotatebox{-270}{\makebox(0,0){\strut{} \huge $E$}}}%
        \put(4006,50){\makebox(0,0){\strut{} \huge $N$}}%
      \put(5006,4400){\makebox(0,0){\strut{} \huge $\gamma=0.4$}}%
      \put(6406,4200){\makebox(0,0){\strut{} \huge $\bf (a)$}}%
      \put(6000,3400){\makebox(0,0){\strut{} \huge $0.5$}}%
      \put(4000,3400){\makebox(0,0){\strut{} \huge $0.6$}}%
      \put(4000,2650){\makebox(0,0){\strut{} \huge $0.7$}}%
      \put(3000,2050){\makebox(0,0){\strut{} \huge $0.8$}}%
  }%
    \gplgaddtomacro\gplfronttext{%
    }%
    \gplbacktext
    \put(0,0){\includegraphics{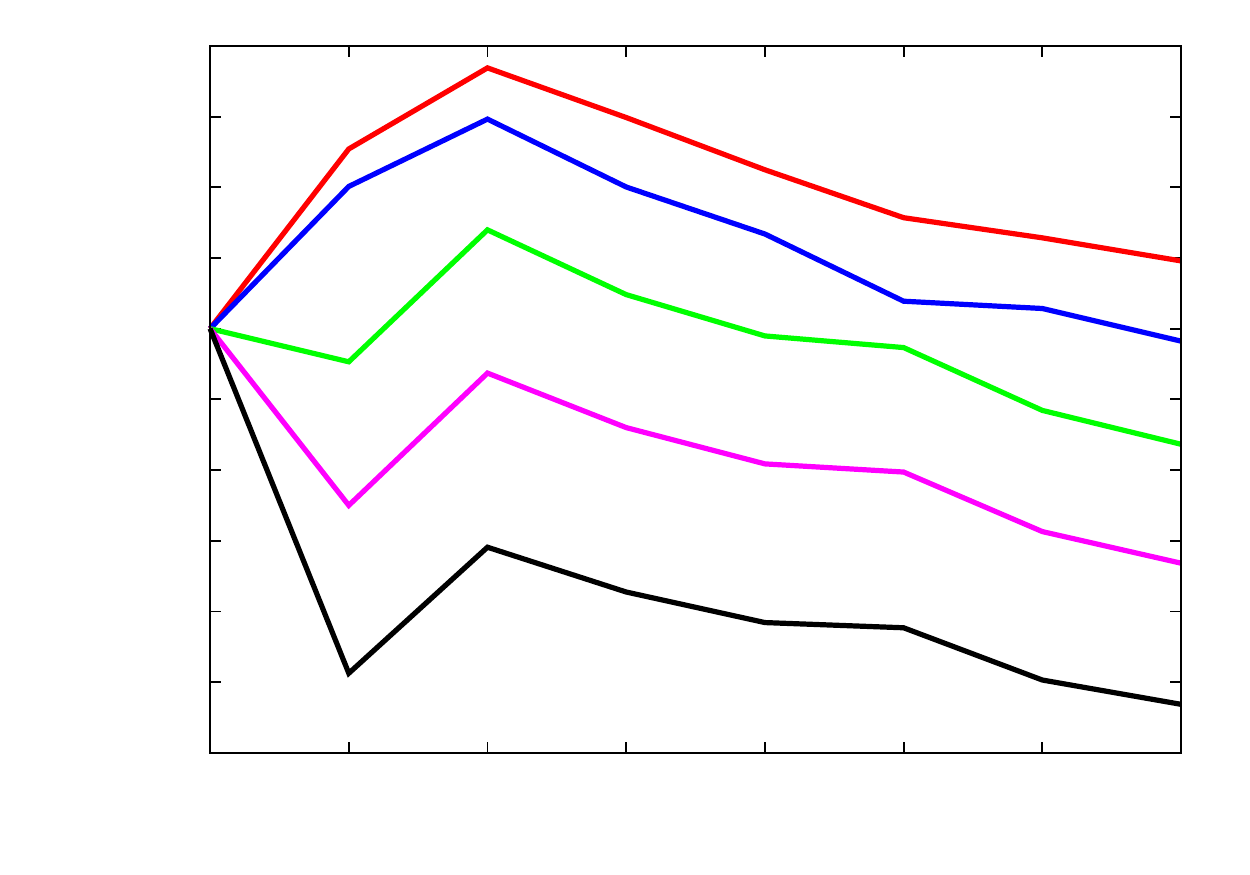}}%
    \gplfronttext
  \end{picture}%
}
\endgroup
\begingroup
  \makeatletter
  \providecommand\color[2][]{%
    \GenericError{(gnuplot) \space\space\space\@spaces}{%
      Package color not loaded in conjunction with
      terminal option `colourtext'%
    }{See the gnuplot documentation for explanation.%
    }{Either use 'blacktext' in gnuplot or load the package
      color.sty in LaTeX.}%
    \renewcommand\color[2][]{}%
  }%
  \providecommand\includegraphics[2][]{%
    \GenericError{(gnuplot) \space\space\space\@spaces}{%
      Package graphicx or graphics not loaded%
    }{See the gnuplot documentation for explanation.%
    }{The gnuplot epslatex terminal needs graphicx.sty or graphics.sty.}%
    \renewcommand\includegraphics[2][]{}%
  }%
  \providecommand\rotatebox[2]{#2}%
  \@ifundefined{ifGPcolor}{%
    \newif\ifGPcolor
    \GPcolortrue
  }{}%
  \@ifundefined{ifGPblacktext}{%
    \newif\ifGPblacktext
    \GPblacktexttrue
  }{}%
  \let\gplgaddtomacro\g@addto@macro
  \gdef\gplbacktext{}%
  \gdef\gplfronttext{}%
  \makeatother
  \ifGPblacktext
    \def\colorrgb#1{}%
    \def\colorgray#1{}%
  \else
    \ifGPcolor
      \def\colorrgb#1{\color[rgb]{#1}}%
      \def\colorgray#1{\color[gray]{#1}}%
      \expandafter\def\csname LTw\endcsname{\color{white}}%
      \expandafter\def\csname LTb\endcsname{\color{black}}%
      \expandafter\def\csname LTa\endcsname{\color{black}}%
      \expandafter\def\csname LT0\endcsname{\color[rgb]{1,0,0}}%
      \expandafter\def\csname LT1\endcsname{\color[rgb]{0,1,0}}%
      \expandafter\def\csname LT2\endcsname{\color[rgb]{0,0,1}}%
      \expandafter\def\csname LT3\endcsname{\color[rgb]{1,0,1}}%
      \expandafter\def\csname LT4\endcsname{\color[rgb]{0,1,1}}%
      \expandafter\def\csname LT5\endcsname{\color[rgb]{1,1,0}}%
      \expandafter\def\csname LT6\endcsname{\color[rgb]{0,0,0}}%
      \expandafter\def\csname LT7\endcsname{\color[rgb]{1,0.3,0}}%
      \expandafter\def\csname LT8\endcsname{\color[rgb]{0.5,0.5,0.5}}%
    \else
      \def\colorrgb#1{\color{black}}%
      \def\colorgray#1{\color[gray]{#1}}%
      \expandafter\def\csname LTw\endcsname{\color{white}}%
      \expandafter\def\csname LTb\endcsname{\color{black}}%
      \expandafter\def\csname LTa\endcsname{\color{black}}%
      \expandafter\def\csname LT0\endcsname{\color{black}}%
      \expandafter\def\csname LT1\endcsname{\color{black}}%
      \expandafter\def\csname LT2\endcsname{\color{black}}%
      \expandafter\def\csname LT3\endcsname{\color{black}}%
      \expandafter\def\csname LT4\endcsname{\color{black}}%
      \expandafter\def\csname LT5\endcsname{\color{black}}%
      \expandafter\def\csname LT6\endcsname{\color{black}}%
      \expandafter\def\csname LT7\endcsname{\color{black}}%
      \expandafter\def\csname LT8\endcsname{\color{black}}%
    \fi
  \fi
  \setlength{\unitlength}{0.0500bp}%
 \resizebox{115pt}{84pt}{   \begin{picture}(7200.00,5040.00)%
    \gplgaddtomacro\gplbacktext{%
      \csname LTb\endcsname%
      \put(1078,704){\makebox(0,0)[r]{\strut{} \huge 0.85}}%
      \put(1078,1383){\makebox(0,0)[r]{\strut{} \huge 0.9}}%
      \put(1078,2061){\makebox(0,0)[r]{\strut{} \huge 0.95}}%
      \put(1078,2740){\makebox(0,0)[r]{\strut{} \huge 1}}%
      \put(1078,3418){\makebox(0,0)[r]{\strut{} \huge 1.05}}%
      \put(1078,4097){\makebox(0,0)[r]{\strut{} \huge 1.1}}%
      \put(1078,4775){\makebox(0,0)[r]{\strut{} \huge 1.15}}%
      \put(1210,284){\makebox(0,0){\strut{} \huge 1}}%
      \put(2009,284){\makebox(0,0){\strut{} \huge 2}}%
      \put(2808,284){\makebox(0,0){\strut{} \huge 3}}%
      \put(3607,284){\makebox(0,0){\strut{} \huge 4}}%
      \put(4406,284){\makebox(0,0){\strut{} \huge 5}}%
      \put(5205,284){\makebox(0,0){\strut{} \huge 6}}%
      \put(6004,284){\makebox(0,0){\strut{} \huge 7}}%
      \put(6803,284){\makebox(0,0){\strut{} \huge 8}}%
      \put(176,2739){\rotatebox{-270}{\makebox(0,0){\strut{} \huge $E$}}}%
      \put(4006,50){\makebox(0,0){\strut{} \huge $N$}}%
      \put(5006,4400){\makebox(0,0){\strut{} \huge $e=0.2$}}%
      \put(6406,4200){\makebox(0,0){\strut{} \huge $\bf (b)$}}%
       \put(3206,3400){\makebox(0,0){\strut{} \huge $0.4$}}%
       \put(5406,2300){\makebox(0,0){\strut{} \huge $0.6$}}%
      \put(4006,2000){\makebox(0,0){\strut{} \huge $0.8$}}%
       \put(2506,1400){\makebox(0,0){\strut{} \huge $1.0$}}%
   }%
    \gplgaddtomacro\gplfronttext{%
    }%
    \gplbacktext
    \put(0,0){\includegraphics{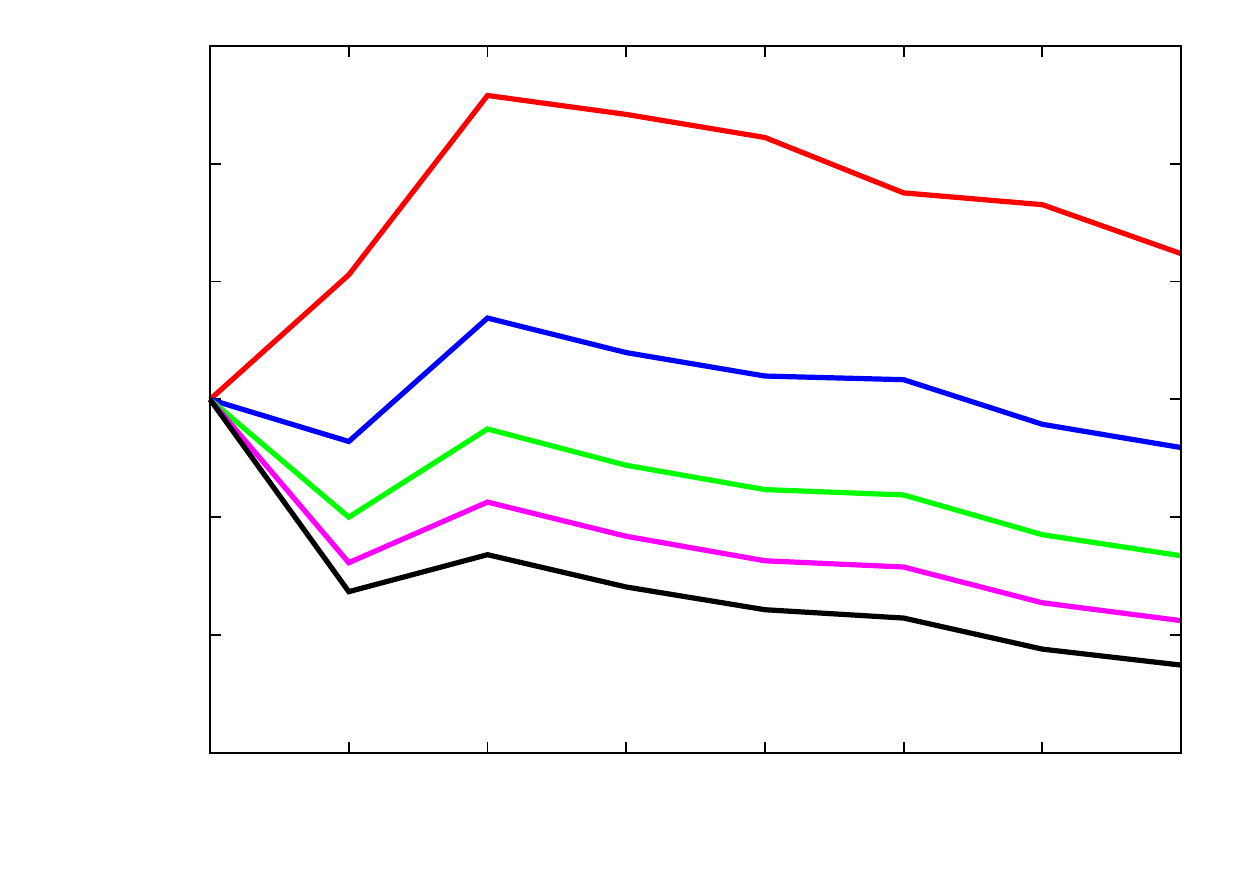}}%
    \gplfronttext
  \end{picture}%
}
\endgroup
\hss}
\vspace{-0.65cm}
\hbox to \linewidth{ \hss 
\begingroup
  \makeatletter
  \providecommand\color[2][]{%
    \GenericError{(gnuplot) \space\space\space\@spaces}{%
      Package color not loaded in conjunction with
      terminal option `colourtext'%
    }{See the gnuplot documentation for explanation.%
    }{Either use 'blacktext' in gnuplot or load the package
      color.sty in LaTeX.}%
    \renewcommand\color[2][]{}%
  }%
  \providecommand\includegraphics[2][]{%
    \GenericError{(gnuplot) \space\space\space\@spaces}{%
      Package graphicx or graphics not loaded%
    }{See the gnuplot documentation for explanation.%
    }{The gnuplot epslatex terminal needs graphicx.sty or graphics.sty.}%
    \renewcommand\includegraphics[2][]{}%
  }%
  \providecommand\rotatebox[2]{#2}%
  \@ifundefined{ifGPcolor}{%
    \newif\ifGPcolor
    \GPcolortrue
  }{}%
  \@ifundefined{ifGPblacktext}{%
    \newif\ifGPblacktext
    \GPblacktexttrue
  }{}%
  \let\gplgaddtomacro\g@addto@macro
  \gdef\gplbacktext{}%
  \gdef\gplfronttext{}%
  \makeatother
  \ifGPblacktext
    \def\colorrgb#1{}%
    \def\colorgray#1{}%
  \else
    \ifGPcolor
      \def\colorrgb#1{\color[rgb]{#1}}%
      \def\colorgray#1{\color[gray]{#1}}%
      \expandafter\def\csname LTw\endcsname{\color{white}}%
      \expandafter\def\csname LTb\endcsname{\color{black}}%
      \expandafter\def\csname LTa\endcsname{\color{black}}%
      \expandafter\def\csname LT0\endcsname{\color[rgb]{1,0,0}}%
      \expandafter\def\csname LT1\endcsname{\color[rgb]{0,1,0}}%
      \expandafter\def\csname LT2\endcsname{\color[rgb]{0,0,1}}%
      \expandafter\def\csname LT3\endcsname{\color[rgb]{1,0,1}}%
      \expandafter\def\csname LT4\endcsname{\color[rgb]{0,1,1}}%
      \expandafter\def\csname LT5\endcsname{\color[rgb]{1,1,0}}%
      \expandafter\def\csname LT6\endcsname{\color[rgb]{0,0,0}}%
      \expandafter\def\csname LT7\endcsname{\color[rgb]{1,0.3,0}}%
      \expandafter\def\csname LT8\endcsname{\color[rgb]{0.5,0.5,0.5}}%
    \else
      \def\colorrgb#1{\color{black}}%
      \def\colorgray#1{\color[gray]{#1}}%
      \expandafter\def\csname LTw\endcsname{\color{white}}%
      \expandafter\def\csname LTb\endcsname{\color{black}}%
      \expandafter\def\csname LTa\endcsname{\color{black}}%
      \expandafter\def\csname LT0\endcsname{\color{black}}%
      \expandafter\def\csname LT1\endcsname{\color{black}}%
      \expandafter\def\csname LT2\endcsname{\color{black}}%
      \expandafter\def\csname LT3\endcsname{\color{black}}%
      \expandafter\def\csname LT4\endcsname{\color{black}}%
      \expandafter\def\csname LT5\endcsname{\color{black}}%
      \expandafter\def\csname LT6\endcsname{\color{black}}%
      \expandafter\def\csname LT7\endcsname{\color{black}}%
      \expandafter\def\csname LT8\endcsname{\color{black}}%
    \fi
  \fi
  \setlength{\unitlength}{0.0500bp}%
\resizebox{230pt}{168pt}{   
  \begin{picture}(7200.00,5040.00)%
  \gplgaddtomacro\gplbacktext{%
    }%
    \gplgaddtomacro\gplfronttext{%
      \csname LTb\endcsname%
      \put(1170,800){\makebox(0,0){\strut{} \large 0.1}}%
      \put(1864,800){\makebox(0,0){\strut{} \large 0.2}}%
      \put(2559,800){\makebox(0,0){\strut{} \large 0.3}}%
      \put(3253,800){\makebox(0,0){\strut{} \large 0.4}}%
      \put(3947,800){\makebox(0,0){\strut{} \large 0.5}}%
      \put(4641,800){\makebox(0,0){\strut{} \large 0.6}}%
      \put(5336,800){\makebox(0,0){\strut{} \large 0.7}}%
      \put(6030,800){\makebox(0,0){\strut{} \large 0.8}}%
      \put(3600,470){\makebox(0,0){\strut{} \large $\gamma$}}%
      \put(998,1086){\makebox(0,0)[r]{\strut{} \large 0.2}}%
      \put(998,1472){\makebox(0,0)[r]{\strut{} \large 0.3}}%
      \put(998,1858){\makebox(0,0)[r]{\strut{} \large 0.4}}%
      \put(998,2244){\makebox(0,0)[r]{\strut{} \large 0.5}}%
      \put(998,2630){\makebox(0,0)[r]{\strut{} \large 0.6}}%
      \put(998,3016){\makebox(0,0)[r]{\strut{} \large 0.7}}%
      \put(998,3402){\makebox(0,0)[r]{\strut{} \large 0.8}}%
      \put(998,3788){\makebox(0,0)[r]{\strut{} \large 0.9}}%
      \put(998,4174){\makebox(0,0)[r]{\strut{} \large 1}}%
      \put(404,2630){\rotatebox{-270}{\makebox(0,0){\strut{} \large $e$}}}%
      \put(6500,4000){\makebox(0,0){\strut{} \huge $\bf (c)$}}%
    }%
    \gplbacktext
    \put(0,0){\includegraphics{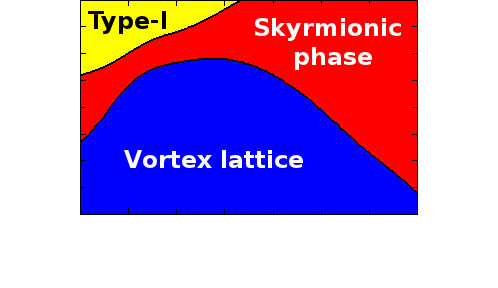}}%
    \gplfronttext
  \end{picture}%
}
\endgroup
 \hss}
\vspace{-0.65cm}
\caption{
(Color on-line) -- 
Upper panels show the dependence of the energy per flux quantum for skyrmions of different topological charges
$\Q$ (values are given in the units of the energy of one integer flux vortex). The  $N=1$ point at the origin
corresponds to an ordinary vortex solution. Panel $\bf (a)$ shows calculations corresponding to different
$\gamma$ for fixed $e=0.6$, while $\bf (b)$ displays how the energy per flux quantum changes with $e$ and $N$
for fixed anisotropy parameter $\gamma=0.7$. 
The $\Q=2$ skyrmions are usually less energetically expensive than the $\Q=3$. This is because the $\Q=2$
skyrmions can be better aligned with the underlying anisotropies, than the $\Q=3$ skyrmions. 
\\
The lower panel displays the \emph{phase diagram}, calculated using energy characteristics of $\Q=5$ 
skyrmions. The different colors refer to different physical properties. The type-I region is shown by yellow 
shade. The lower part of the phase diagram shows regions where skyrmions (red) or vortex lattices (blue) 
form in applied external field. The phase diagram retains similar structure  in calculations with different 
topological charges. With the increasing of the skyrmionic charge $\Q$, the region where skyrmions are 
energetically preferred over vortex lattices grows slightly. These results apply either for two-dimensional 
systems or three dimensional systems with translational invariance along $z$-axis. In the latter case the 
energy should be understood as the energy per unit length of a skyrmion line  (i.e. a skyrmion texture in 
$xy$ plane which is invariant under translation along $z$-axis). 
The discretization errors can be estimated by computing the total magnetic  flux and comparing it to the
exact value which follows from the quantization condition $2\pi N/e$. This gives the relative accuracy on
the flux to be around $10^{-5}$. From that, the accuracy on the energy is estimated to be at least three 
order of magnitude smaller than the energy difference between skyrmions and vortices.
}
\label{Skyrmion-energy1}
\end{figure}

The main result of this work is that skyrmions of the above type (and thus half-quantum vortices) can be 
\emph{less energetic} than integer-flux ordinary vortices and
\emph{thermodynamically stable}, in the chiral $p$-wave superconductors. The critical external magnetic field
${H}_{c1}$ for the formation of a flux-carrying topological defect is determined by the condition where the Gibbs free
energy  $G=E_d-2\int {\bf B}\cdot {\bf H}_e\,\, \dd x\dd y$ becomes negative. Here $E_d$ and ${\bf B}$ are
the energy and magnetic field of the defect. ${\bf H}_e$ denotes the applied field. Thus $H_{c1}= E_d/2\Phi$ 
where $\Phi$ is the magnetic flux produced by the defect. The defects are thermodynamically  stable if  the  critical
external magnetic field's energy density $H_{c1}^2$ is smaller than the condensation energy.
We investigated the energy dependence of the skyrmions on the number of enclosed
flux quanta $N$. The energy of an integer flux vortex  is used as a reference point. As shown
in \Figref{Skyrmion-energy1} panels $\bf(a)$ and $\bf(b)$, for low $N$, the energy depends non-monotonically
on $N$. This is because the preferred symmetry of small $N$ configurations in some cases is in strong conflict
with the anisotropies of the model. In the large-$N$ limit the energy per flux quantum gradually tends to some
value. The main point here is that the energy per flux quantum for skyrmions is 
in certain cases smaller than that of vortices. This signals
instability of vortex lattices with respect to skyrmion formation.

\begin{table}[!htb]
\begin{center}
\begin{tabular}{|p{0.12\linewidth}||p{0.15\linewidth}|p{0.15\linewidth}|p{0.15\linewidth}|p{0.15\linewidth}|p{
0.15\linewidth}|}
\hline
 { }		& {  $E_\mathrm{total}$ } & { $E_\mathrm{grad}$}  & {$E_\mathrm{pot}$} 
& {$E_{\Ztwo}$} & {$E_\mathrm{mag}$} 
 \\ \hline \hline
 {Vortex}& {19.7759} & {10.7518}  	& {-12.0190} 	& {16.5195}	&{4.5235}  
\\ \hline 
 {Skyrm.}& {18.9004} & {8.10522}	& {-12.2301} 	& {17.6336}	&{5.3916}  
\\ \hline \hline 
 {Vortex}& {32.1684} & {19.3227}  	& {-19.0381} 	& {25.4445}	&{6.4392}  
\\ \hline 
 {Skyrm.}& {37.6456} & {16.2529}  	& {-22.1474} 	& {32.8582}	&{10.6818}  
\\ \hline \hline 
\end{tabular}
\caption{
Different contributions to the  skyrmion energy \emph{per flux quantum}. $\Q=5$ skyrmions are considered in
this example. The results are compared with the contributions to the energy of a single vortex (which
determines the lower bound on vortex lattice energy near  the first critical magnetic field $H_{c1}$).
 The gradient contribution $E_\mathrm{grad}$ is given by the integrated \Eqref{gradientEnergy}, the magnetic
energy $E_\mathrm{mag}$ by \Eqref{magneticEnergy}. The potential energy $E_\mathrm{pot}$ is
\Eqref{potentialEnergy} and $E_{\Ztwo}$ is \Eqref{breakEnergy}. 
First block, for which $\gamma=0.8$ and $e=0.4$, corresponds to the state where skyrmions are thermodynamically
stable but vortex lattices are not. 
Second block is for $\gamma=0.6$ and $e=0.2$. It corresponds to a regime with standard Abrikosov vortex
lattice. Here the skyrmions are local minima of the free energy functional. They are more expensive than
vortices but, if formed, they are protected against decay by a finite energy barrier. 
In the second example the win in the kinetic energy is too small to overcome extra energy cost associated with
domain wall formation and magnetic energy.
}
\label{table1}
\end{center}
\end{table}

Next, the thermodynamical stability of skyrmions is investigated. Results for $N=5$ quanta are reported
as a characteristic example, in \Figref{Skyrmion-energy1}$~\bf (c)$. We find that there are three regimes 
on the resulting phase diagram. When the penetration length is large (\ie low $e$), the system shows usual the type-II
superconductivity.  When the penetration length is small, the system is a type-I superconductor. For intermediate
values of the penetration length, depending on the underlying anisotropies $\gamma$, the external field
produces skyrmions rather than vortex lattices. 
To understand the instability of vortex lattices with respect to skyrmion formation, different contributions
to energy are investigated in Table~\ref{table1}. In the skyrmionic state, vortex lattice decay into skyrmions 
is driven by a win in gradient and potential energies although there is a loss in magnetic energy as well as the 
extra cost of producing a domain wall.

The skyrmions  we find are are structurally different from skyrmions discussed in other kinds of
superconductors \cite{Knigavko.Rosenstein:99,*Knigavko.Rosenstein.ea:99,*Rosenstein.Shapiro.ea:03,
*Li.Toner.ea:09,*Li.Toner.ea:07}  because of the different symmetry
of the model. Another principal difference is the nature of the skyrmionic state,  
namely Ref.~\onlinecite{Knigavko.Rosenstein:99,*Knigavko.Rosenstein.ea:99,*Rosenstein.Shapiro.ea:03,
*Li.Toner.ea:09,*Li.Toner.ea:07} proposed models where there are only skyrmionic solutions 
carrying two flux quanta. The latter forming stable lattices. In contrast, the model  we consider supports
skyrmions with any integer value of topological charge.
Importantly, the energy per flux quantum here is a sublinear function of the topological charge, which
prohibits a ground state in the form of a lattice  of the simplest skyrmions envisaged
in Ref.~\onlinecite{Knigavko.Rosenstein:99,*Knigavko.Rosenstein.ea:99,*Rosenstein.Shapiro.ea:03,
*Li.Toner.ea:09,*Li.Toner.ea:07}. Instead our model
predicts more complicated high-topological-charge skyrmionic structures.
Also in type-II regime our model predicts metastable states of coexisting vortices and skyrmions.


In conclusion we have shown that the phase diagram of chiral $p$-wave superconductors has a 
thermodynamically stable skyrmionic phase between type-I and the usual type-II regimes. This is 
despite the fact that the model has $\Uone\times\Ztwo$ broken symmetry where naive symmetry 
arguments would rule out skyrmionic excitations. In the skyrmionic phase,  the long sought-after 
half-quantum vortices acquire thermodynamic stability. These objects can be detected with surface probes 
through their characteristic profile of magnetic field. The phase transition into a skyrmionic state 
should be first order, because the energy per flux quantum is decreasing with the skyrmionic 
topological charge.

The possibly chiral superconductor \SRO which is frequently described by the model \Eqref{freeEnergy} 
may have a penetration length which is slightly too large to fall into the skyrmionic phase. However in
this case, the model predicts metastable skyrmionic excitations (which are slightly more 
energetic than vortices). Recently sporadic formation of objects with multiple flux 
quanta were reported in Fig.~2 of Ref.~\onlinecite{moler}. Higher resolution scans of the magnetic
field profile could confirm or rule out if the observed objects are skyrmions.
Another scenario for flux clustering in this material is type-1.5 superconductivity which 
can arise if to take into account its multi-band nature \cite{Garaud.Agterberg.ea:12}.

\begin{acknowledgments}
We thank D.~F.~Agterberg and  J.~Carlstr\"om for discussions.
The work is supported by the Swedish Research Council, and by the Knut and Alice Wallenberg Foundation through
the Royal Swedish Academy of Sciences fellowship and by NSF CAREER Award Nos. DMR-0955902. 
The computations were performed on resources provided by the Swedish National Infrastructure 
for Computing (SNIC) at National Supercomputer Center at Linkoping, Sweden.
\end{acknowledgments}

%

\clearpage
\appendix
\renewcommand{\theequation}{\Alph{section}.\arabic{equation}}

\section{Technical details} \label{Appendix}

Functional variation of the free energy  \Eqref{freeEnergy} with respect to the fields provides 
Euler-Lagrange equations of motion. In particular, variation with respect to the gauge 
field defines the total supercurrent 
\Equation{Supercurrent1}{
   \bs J\equiv\bs J^\oo + \bs J^\ot \,.
}
The contribution of each component there, is given by 
\Align{Supercurrent2}{
   J^\oo_x & = \frac{e}{2}\Im\left[\psi_1^*(D_x\psi_1+\gamma D_y\psi_2) \right] 		
\nonumber \\
   J^\oo_y & = \frac{e\gamma}{2}\Im\left[\psi_1^*(D_y\psi_1+ D_x\psi_2) \right] 		
\nonumber \\
   J^\ot_x & = \frac{e\gamma}{2}\Im\left[\psi_2^*(D_x\psi_2+ D_y\psi_1)  \right] 		
\nonumber \\
   J^\ot_y & = \frac{e}{2}\Im\left[\psi_2^*(D_y\psi_2+\gamma D_x\psi_1)  \right] 			\,.
}
The contributions $|\bs J^\oo|$ and $|\bs J^\ot|$ are displayed for the different solutions. 
Each contribution to the supercurrent reflects the underlying anisotropies due to 
the complicated gradient terms \Eqref{gradientEnergy}.

\subsection*{Simple quantum vortex solutions} 
\begin{figure}[!htb]
  \hbox to \linewidth{ \hss
  \includegraphics[width=\linewidth]{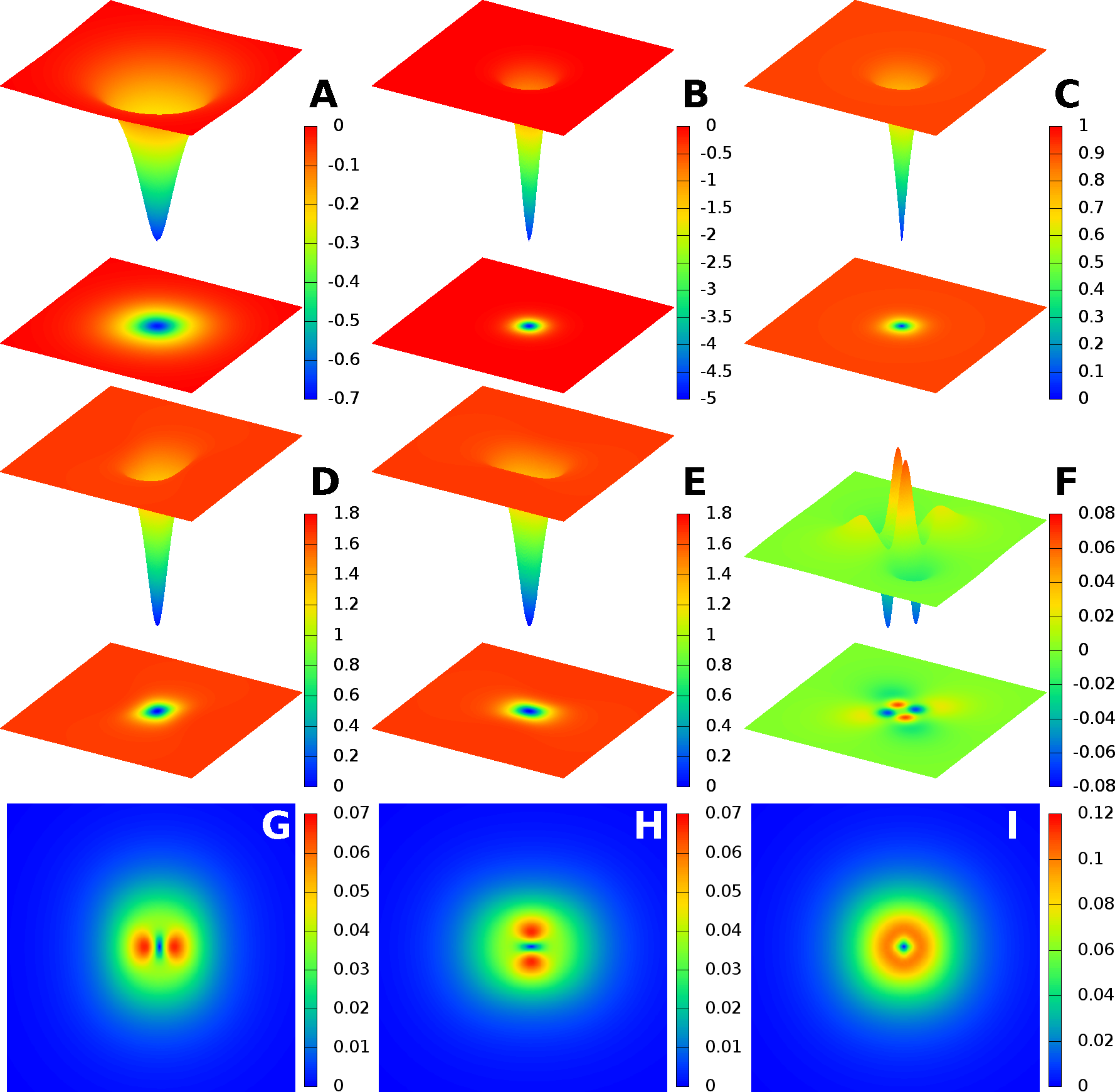}
 \hss}
\caption{
(Color on-line) -- 
A single vortex solution, when $e=0.2$ and the parameter $\gamma=0.6$.  
Displayed quantities are, magnetic flux ($\bf A$), the (inverted) energy density ($\bf B$) and  the sine of
the phase difference $\sin(\varphi_{2}-\varphi_{1})$ ($\bf C$). On the second line, the densities
of superconducting order parameter components  $|\psi_1|^2$ ($\bf D$), $|\psi_2|^2$ ($\bf E$), and
the `doubled phase difference' $\Im(\psi_1^{*\ 2}\psi_2^2)$ ($\bf F$). Panels ($\bf G$) and (\resp $\bf H$) on
the third line, are contribution $|{\bs J}^\oo|$ (\resp $|{\bs J}^\ot|$) of $\psi_1$ (\resp $\psi_2$)
component 
to supercurrent \Eqref{Supercurrent2}. The last panel ($\bf I$) shows the total supercurrent
\Eqref{Supercurrent1}.
The densities ($\bf D$ and $\bf E$) and the supercurrents ($\bf G$ and $\bf H$) 
of the different components of an integer quantum vortex are very anisotropic. 
The anisotropies are less perceptible when considering the magnetic field ($\bf A$) or 
the energy ($\bf B$).
The main difference with the skyrmions can be seen from panel $\bf C$. Indeed the phase difference 
goes from zero at the vortex core to $\pi/2$ faraway, instead of $-\pi/2$ to $\pi/2$ for a skyrmions. 
}
\label{Fig:Vortex}
\end{figure}
%
In the main part of the text, skyrmionic excitations 
 are discussed. 
For a comparison, we display in \Figref{Fig:Vortex} the solution for a usual
integer quantum vortex solution in the model \Eqref{freeEnergy}.
Because of the anisotropies \Eqref{gradientEnergy} of the theory, the single 
quantum vortex is also non-axially symmetric. In contrast to vortices, skyrmions have a non-zero skyrmionic topological charge. 
Moreover, visual inspection of phase difference $\varphi_{2}-\varphi_{1}$ of the skyrmions 
provides further arguments of the qualitative difference from vortices. Indeed, for skyrmions, 
the phase difference $\varphi_{2}-\varphi_{1}$ covers the range $\left[-\pi/2,\pi/2\right]$. 
Thus it links both inequivalent ground states. Phase difference for integer vortex ranges only 
$\left[0,\pi/2\right]$.

\subsection*{\texorpdfstring{$\psi_\pm$ parametrization of the condensates}
{psi-pm parametrization of the condensates}
} 
\begin{figure}[!htb]
  \hbox to \linewidth{ \hss
  \includegraphics[width=\linewidth]{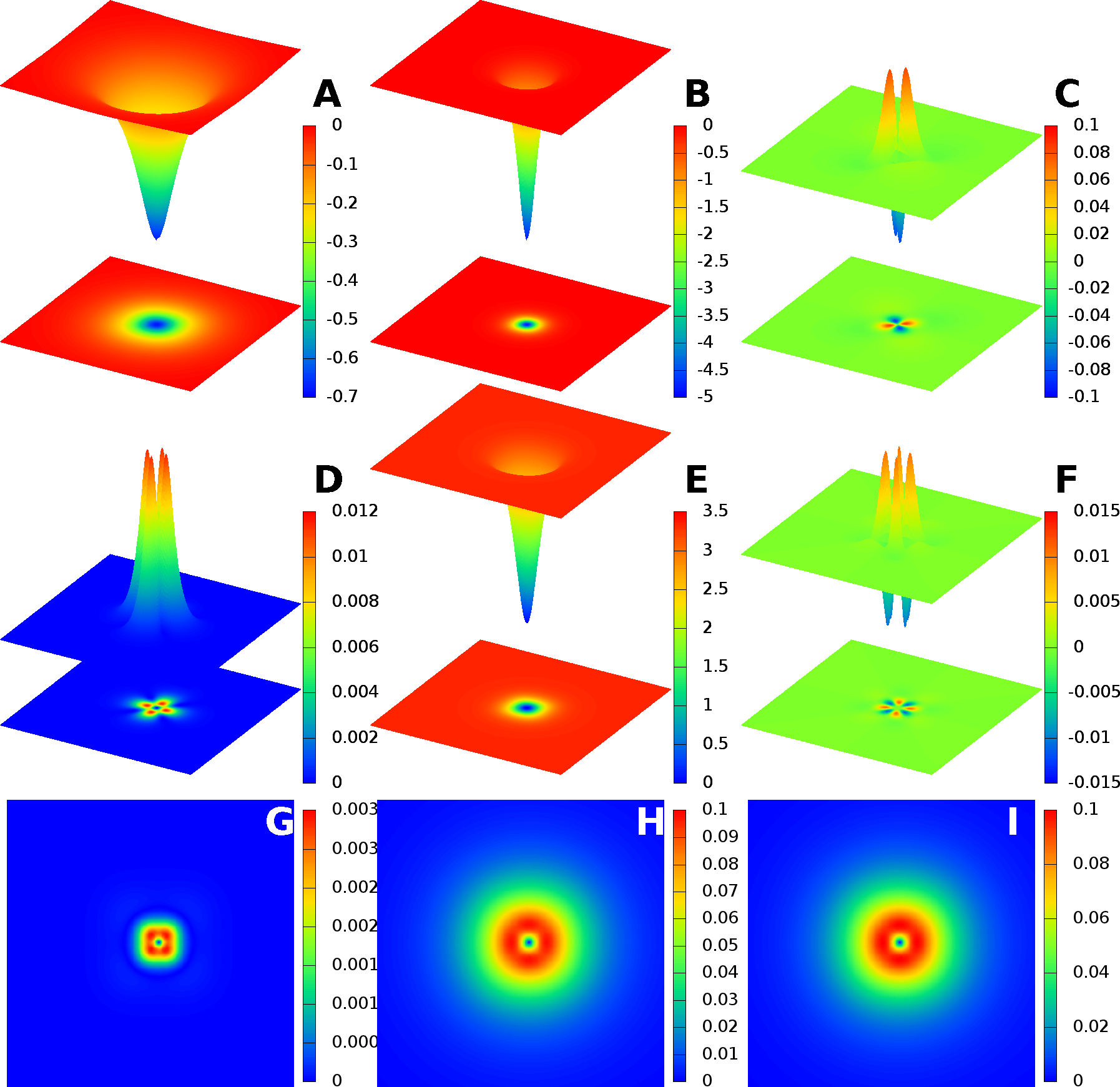}
 \hss}
\caption{
(Color on-line) -- 
A single vortex solution, for the same parameters as in \Figref{Fig:Vortex}.  
Quantities displayed now are using the $\psi_\pm$ parametrization.
Panel ($\bf A$) is the magnetic flux, and ($\bf B$) is the (inverted) energy density. 
The sine of the phase difference $\sin(\varphi_{+}-\varphi_{-})$ ($\bf C$). On the second line, the densities
of superconducting order parameter components  $|\psi_+|^2$ ($\bf D$), $|\psi_-|^2$ ($\bf E$), and
the `doubled phase difference' $\Im(\psi_+^{*\ 2}\psi_-^2)$ ($\bf F$). Panels ($\bf G$) and (\resp $\bf H$) on
the third line, are contribution $|{\bs J}^\om|$ (\resp $|{\bs J}^\op|$) of $\psi_+$ (\resp $\psi_-$)
component 
to supercurrent \Eqref{Supercurrent4}. The last panel ($\bf I$) shows the total supercurrent
\Eqref{Supercurrent3}.
}
\label{Fig:Vortex-PM}
\end{figure}
%

Another parametrization of the condensate is sometimes used in literature. Namely  
combinations 
\Equation{Reparametrization}{
   \psi_\pm\equiv\frac{\psi_1\pm\psi_2}{\sqrt{2}}
}
are used instead of $\psi_{1,2}$. The free energy functional \Eqref{freeEnergy} can be rewritten within this
different representation (a special care is there needed in the redefinition of the parameters).
Let the total current be
\Equation{Supercurrent3}{
   \bs J\equiv\bs J^\op + \bs J^\om \,.
}
%
\begin{figure}[!htb]
  \hbox to \linewidth{ \hss
  \includegraphics[width=\linewidth]{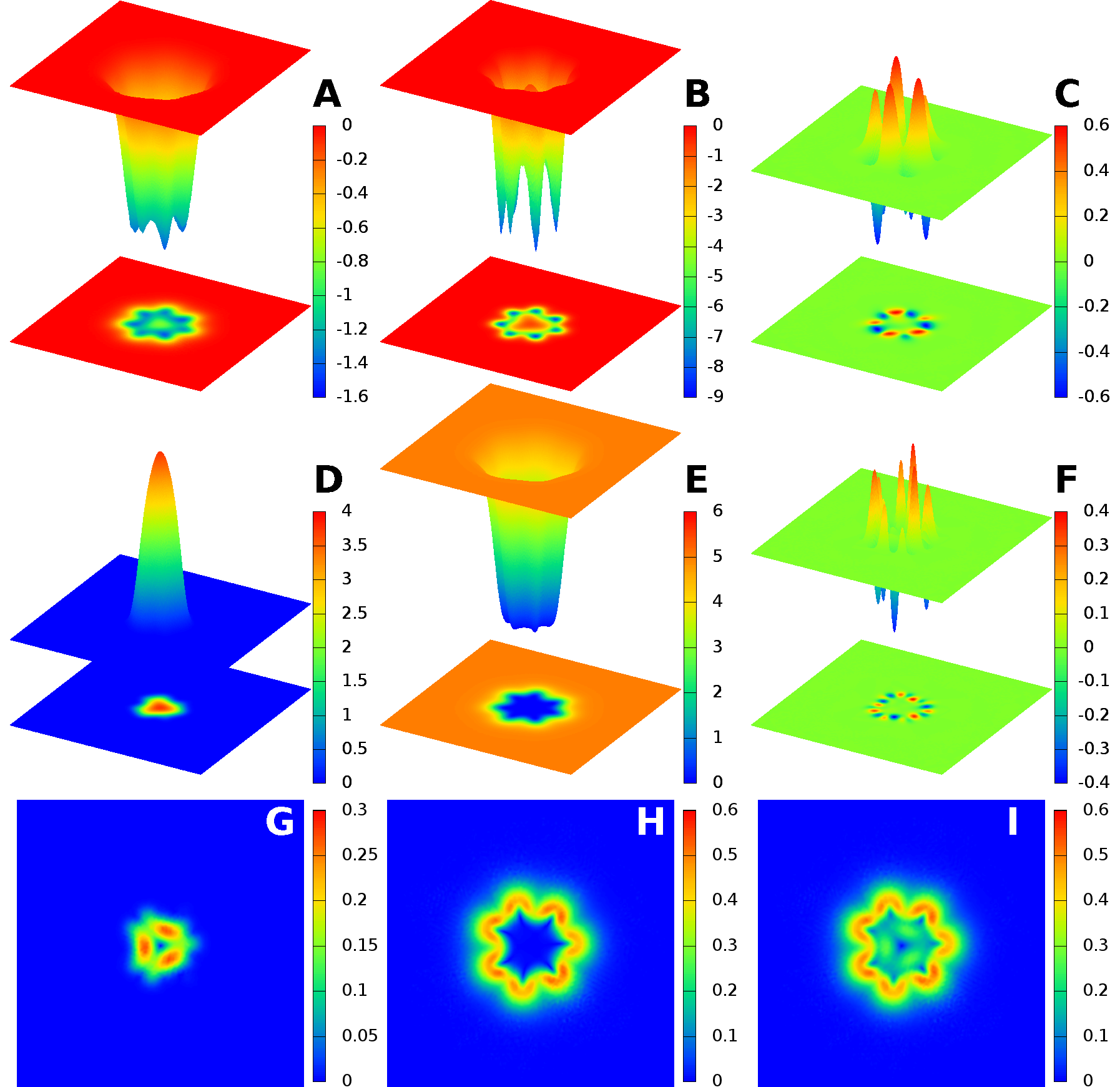}
 \hss}
\caption{
(Color on-line) -- 
The thermodynamically stable skyrmion carrying five flux quanta, 
presented in Fig.~3 of the manuscript, using the $\psi_\pm$ parametrization.
Displayed here quantities are the same as in \Figref{Fig:Vortex-PM}.
}
\label{Fig:Skyrmion1-PM}
\end{figure}
%
The contribution of each component in terms of the $\psi_{1,2}$ parametrization, is given by 
\Align{Supercurrent4}{
   J^\opm_x & =\frac{J^\oo_x+J^\ot_x}{2}	\nonumber \\
&\pm\frac{e}{4}\Re\left[\psi_1^*\gamma(D_x\psi_2+D_y\psi_1)
-\psi_2^*(D_x\psi_1+\gamma D_y\psi_2) \right] 	
\nonumber \\
  J^\opm_y & =\frac{J^\oo_y+J^\ot_y}{2}	\nonumber \\
&\pm\frac{e}{4}\Re\left[\psi_1^*(D_y\psi_2+\gamma D_x\psi_1)
-\psi_2^*\gamma(D_y\psi_1+D_x\psi_2) \right] 	,
 }
where $\bs J^\oo$ and $\bs J^\ot$ are defined in \Eqref{Supercurrent2}. 

In order to compare the features of this new parametrization, \Figref{Fig:Vortex-PM} show the very same
integer flux vortex solution as in \Figref{Fig:Vortex-PM}, but using the parametrization
\Eqref{Reparametrization}. In this new parametrization, the component $\psi_+$ (panel $\bf D$) has zero
ground state density, while $\psi_-$ (panel $\bf E$) recovers to non-zero ground state density far from the
vortex. Both $\psi_+$ and $\psi_-$ components have a coinciding core singularity. Panel $\bf C$, showing the
sine of the phase difference $\sin(\psi_--\psi_+)$ exhibit a non-zero winding around the vortex core. 
This kind of features reproduce the kind of single vortex solution obtained in 
\cite{Heeb.Agterberg:99}. 

%
\begin{figure}[!htb]
  \hbox to \linewidth{ \hss
  \includegraphics[width=\linewidth]{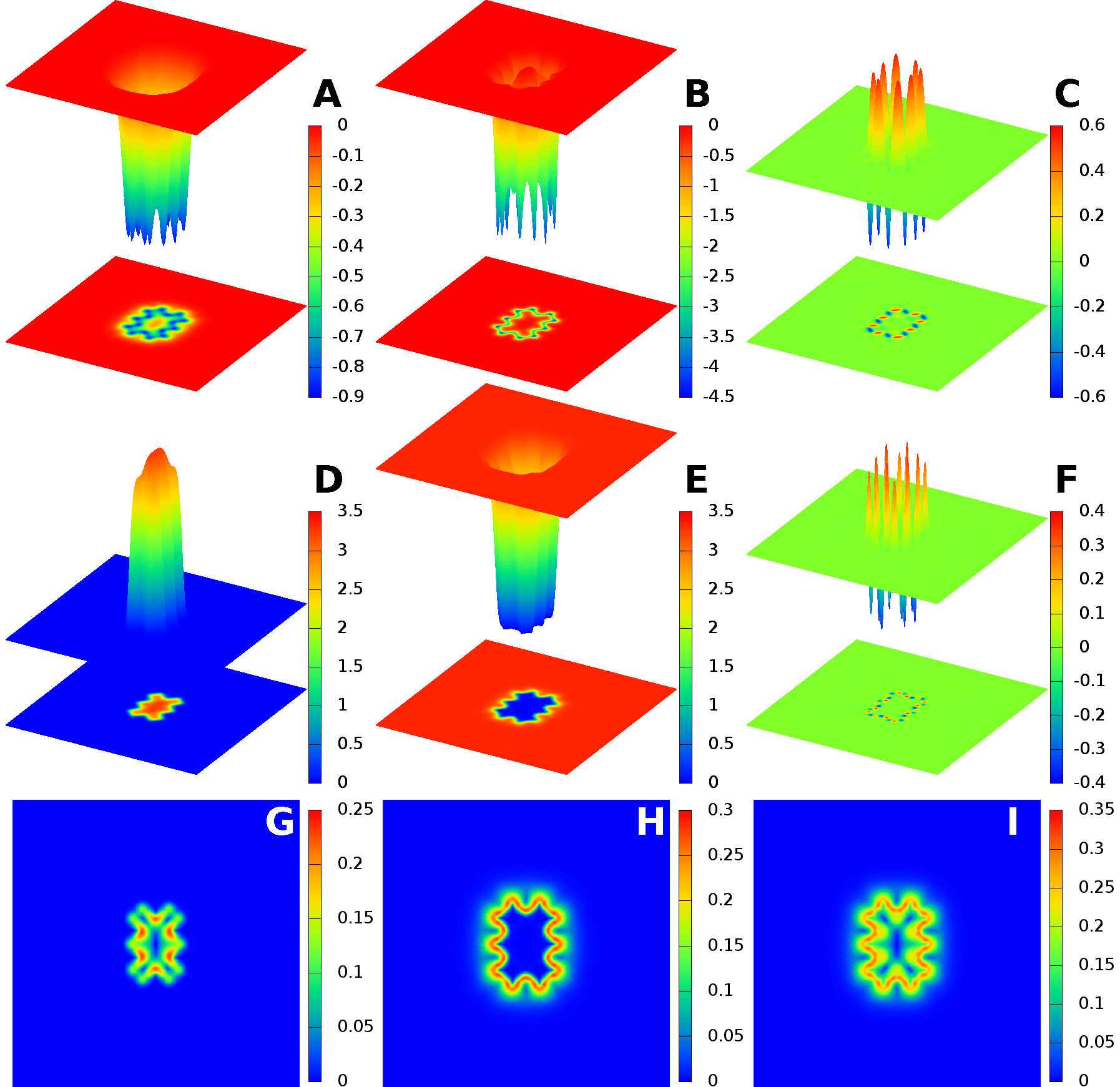}
 \hss}
\caption{
(Color on-line) -- 
The skyrmion carrying eight flux quanta, 
presented in Fig.~4 of the manuscript, using the $\psi_\pm$ parametrization.
Displayed here quantities are the same as in \Figref{Fig:Vortex-PM}.
}
\label{Fig:Skyrmion2-PM}
\end{figure}
%
The skyrmion solutions are displayed within this field parametrization in Figures \ref{Fig:Skyrmion1-PM} 
and \ref{Fig:Skyrmion2-PM}. They correspond to solutions provided in the main part.


\section{Numerical Methods -- Finite element energy minimization} \label{Numerics}

We provide here details about the numerical methods which are used to construct skyrmions 
or vortex solutions of the two-components Ginzburg-Landau model \Eqref{freeEnergy}.
The variational problem is defined for numerical computation using a finite element formulation provided by
the Freefem++ library \cite{Freefem}. Discretization within finite element formulation is done via a
(homogeneous) triangulation over $\Omega$, based on Delaunay-Voronoi algorithm. The domain $\Omega$ is chosen
here to be a disc whose radius is much larger than the vortex/skyrmion size. In most cases the radius of the
disc is  $10$ to $20$ times larger than the size of a single vortex. This guarantees that all the fields reach
their  ground state values at the boundary. This ensures that the topological solitons are not affected by the
boundary. We performed the additional check that solutions are not boundary artifacts  by computing the energy
on the boundary. When the algorithm converges, this quantity is of the order of the numerical accuracy,
which indicates that the solutions do not interact with the boundaries of the numerical domain.
Functions describing the fields are decomposed on a continuous piecewise quadratic basis over each triangle. 
The accuracy of such method is controlled through the number of triangles, (we typically used $3\sim6\times10^4$), 
the order of expansion of the basis on each triangle (P2 elements being 2nd order polynomial basis on each triangle), 
and also the order of the quadrature formula for the integral on the triangles. 

Once the problem is mathematically well defined, a numerical optimization algorithm is used to solve the
variational non-linear problem (\ie to find the minima of $\mathcal{F}$). Here we used a non-linear conjugate
gradient method. The algorithm is iterated until relative variation of the norm of the gradient of the
functional  $\mathcal{F}$ with respect to all degrees of freedom is less than $10^{-6}$.

\end{document}